# Statistically significant observation of and cross sections for a new nuclear reaction channel on $^{197}$Au with bound dineutron escape

IHOR M. KADENKO [1(a)], BARNA BIRÓ[2], ANDRÁS FENYVESI[2]

[1] *International Nuclear Safety Center of Ukraine; Department of Nuclear Physics, Taras Shevchenko National University of Kyiv – St. Volodymyrs`ka, 64/13, 01601, Kyiv, Ukraine*
[2] *Institute for Nuclear Research, Hungarian Academy of Sciences (MTA Atomki), Bem tér 18/c, H-4026 Debrecen, Hungary*

PACS 21.45.Bc  - Two-nucleon system

PACS 25.90.+k  - Other topics in nuclear reactions: specific reactions

PACS 27.10.+h  - A Ö5

**Abstract** – A new nuclear reaction channel on $^{197}$Au with the neutron as a projectile and a bound dineutron ($^2n$) in the output channel is considered based on available experimental observations. The dineutron is assumed to be formed as a particle-satellite, separated from the volume but not from the potential well of $^{196}$Au nucleus. The dineutron was identified by statistically significant radioactivity detection due to decay of $^{196g}$Au nuclei. Cross sections for the $^{197}$Au $(n,^2n)$ $^{196g}$Au reaction are determined as 180 ± 60 $\mu$b and 37 ± 8 $\mu$b for [6.09-6.39] and [6.175-6.455] MeV energy ranges, correspondingly. Possible outcomes of dineutron detection near the surface of deformed nuclei are also raised and discussed.

**Introduction.** The purpose of this letter is discussion of the dineutron as a bound particle, or two-nucleon nucleus, consisting of the two neutrons only without any nucleus charge and formed in the outgoing channel on neutron induced nuclear reaction on $^{197}$Au near the $^{196}$Au nucleus surface. Such a configuration is different from the classical description of nuclear reactions at low energies and described in details in [1]. The dineutron was predicted by A. Migdal to be formed as a bound particle under certain circumstances when an additional bound state appears, not existing in the perturbation theory. This bound state is interpreted as a single particle state for the two neutrons, or bound dineutron, located near the surface of some nucleus, beyond its volume but within its potential well. This bound state corresponds to single-particle level at an additional energy branch, which concludes at $\varepsilon_c \sim 0.4$ MeV. As it is well known, as a minimum, only 0.066 MeV is necessary to bind the two neutrons in the dineutron and this mechanism could make a possible formation of a bound dineutron in the outgoing channel of some nuclear reactions. First observation of this phenomenon was described in [2], but statistical significance of detection of $^{158}$Tb induced activity as a product of the $^{159}$Tb $(n,^2n)$ nuclear reaction was not good enough due to 180 years half-life of $^{158g}$Tb residual nucleus.

Thus, we decided to search for another nucleus with a reasonable half-life to make sure the statistical significance of its decay detection will meet the 5 criteria. Another requirement to select a proper nucleus should be a very high purity of a sample for neutron irradiation to exclude any overlapping due to unexpected interfering nuclear reactions on impurities. As a result, our search procedure resulted in selection of $^{197}$Au as a nucleus-candidate for statistically significant observation of a bound dineutron. Then the presence of the dineutron in the outgoing channel of the $^{197}$Au $(n,^2n)^{196}$Au nuclear reaction would be greatly facilitated when the residual nucleus $^{196g}$Au decays with emission of corresponding gamma-lines, that can then be reliably detected. Finally, our goal was to succeed with dineutron observation in a configuration, similar to presented in fig.1, and to make cross-section estimates for the $^{197}$Au$(n,^2n)^{196g}$Au reaction introducing a new nuclear reaction channel with bound dineutron escape.

**Experimental observations.** - To obtain expected results, golden foil samples were selected for neutron irradiation followed by subsequent gamma-ray counting.

Two irradiation experiments were carried out at Atomki, Hungary. Figure 2 shows the sketch of the arrangement of the irradiations. The quasi-monoenergetic $(d,D)$ neutrons were produced via bombarding a $D_2$-gas target [4] with deuterons accelerated in the MGC-20E cyclotron. The energies of the deuterons were $E_d = 3.459$ MeV and $E_d = 3.523$ MeV. The energy spread of the analysed deuteron beam was $\delta E_d/E_d = 0.1\%$. The diameter of the bombarding deuteron beam was 4

[(a)]E-mail: IMKadenko@univ.kiev.ua

I.M. Kadenko *et al.*

mm and the window of the gas cell was a 5 μm thick Nb foil.

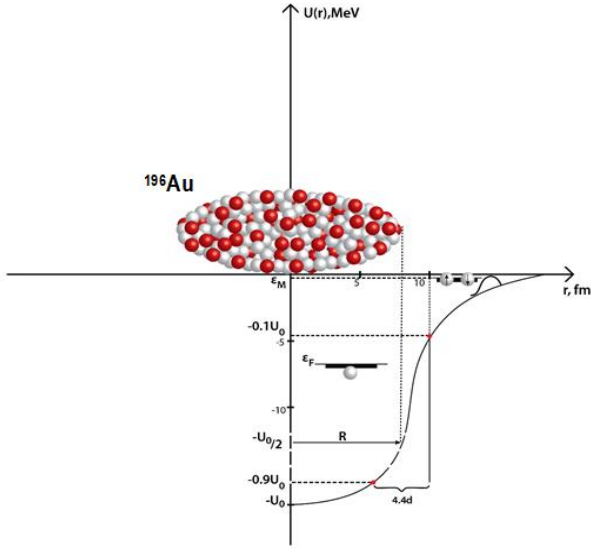

Fig. 1: (Colour on-line) Schematical representation of $^{196}$Au and the dineutron in the outgoing channel of the $^{197}$Au$(n,^2n)^{196g}$Au reaction.

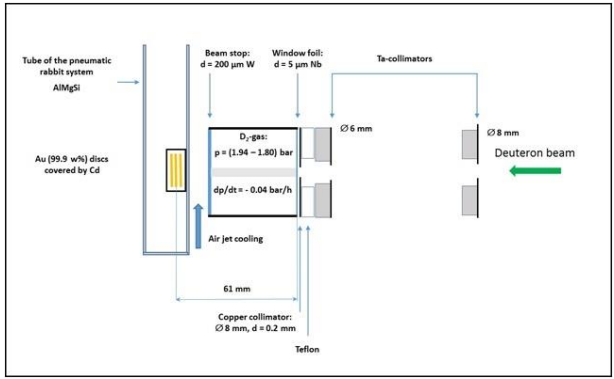

Fig. 2: (Colour on-line) Sketch of the arrangements for irradiations.

The pressure of the $D_2$-gas was kept in the $p = (1.94 \div 1.80)$ bar range and a pressure loss $dp/dt = -0.04$ bar/hour was compensated by re-filling the gas cell periodically every 1.5 hours. The beam stop of the gas cell was a 200 μm thick tungsten plate. The distance between the window foil and the beam stop and, thus, the length of the bombarded gas volume, was 39.4 mm. The beam stop was cooled by an air jet. Typically, $I_{coll-1} = 400$ nA current was measured on the dia. 8 mm collimator and the $I_{coll-2} \leq 50$ nA current on the dia. 6 mm collimator to be within the range of the leakage current of the system.

The samples used for the experiment done at $E_d = 3.459$ MeV deuteron energy were cut from a 100 μm thick 99.9 w% Au sheet of chemical purity. Three pieces of this purity and 100 μm thick and 1.3 cm diameter foil samples (Au-1, Au-2 and Au-3) were put in a closed cadmium cover of 0.51 mm wall thickness to reduce intensity of 411.8 keV gamma line due to $^{198}$Au decay because of scattered neutrons and possible yield enhancement for $^{197}$Au $(n,\gamma)$ $^{198}$Au nuclear reaction. Such sample configuration ensured us to avoid a situation when gamma-peaks of our interest with about 333 and 356 keV energies will not be hidden in the Compton distribution due to higher energy gamma-peaks. The internal diameter of the cover was 13 mm. The 99.9 w% chemical purity of the Au material was checked by Secondary Ion Mass Spectrometry (SIMS) at Atomki. Presence of Hg (0.1 w%) with natural isotopic composition and minor Bi (< 0.01 w%) were detected as impurities. The number of $^{196}$Hg atoms in each of the Au-1, Au-2 and Au-3 samples was $N_{Hg-196} = 1.154 \cdot 10^{15} \pm 2\%$.

The quasi-monoenergetic peak of the spectrum of the $(d,D)$ neutrons covered the $[6.09 \div 6.39]$ MeV neutron energy range. The average current of the deuteron beam was $I_d = 1.95$ μA and the total amount of the charge of deuterons delivered to the target was $Q_d = 2.73 \cdot 10^{-2}$ C. The neutron fluence was $\Phi_{total} = 1.33 \cdot 10^{11}$ cm$^{-2}$ delivered at $\langle \phi_{neutron} \rangle = d\langle \Phi_n \rangle/dt = 9.5 \cdot 10^6$ cm$^{-2}$s$^{-1}$ average neutron fluence rate. For the experiment done at $E_d = 3.523$ MeV deuteron energy Au foils of 12.2 mm diameter and 150 μm thickness (Au-I and Au-II) and one more Au foil of the same diameter and 200 μm thickness (Au-III) were stacked and covered by 1 mm thick cadmium shielding. Then the stack was irradiated with quasi-monoenergetic neutrons for 7 809 s. The energy spectrum of $(d,D)$ neutrons covered the $[6.175-6.455]$ MeV energy range. The average current of the deuteron beam was $I_d = 1.92$ μA and the total amount of the charge of deuterons delivered to the target was $Q_d = 1.51 \cdot 10^{-2}$ C. The neutron fluence was $\Phi_{total} = 7.5 \cdot 10^{10}$ cm$^{-2}$ and it was delivered at $\langle \phi_{neutron} \rangle = d\langle \Phi_n \rangle/dt = 9.6 \cdot 10^6$ cm$^{-2}$s$^{-1}$ average neutron fluence rate to the samples.

The NeuSDesc code [5] was used for calculation of the spectra of the $(d,D)$ neutrons and the neutron fluence rates for the positions of the geometry centres of the foil stacks. The neutron spectra obtained are shown in fig. 3 that are far below the $E_{th\ (n,2n)} = 8.11372$ MeV threshold energy of the $^{197}$Au$(n,2n)^{196}$Au nuclear reaction.

The induced activities of the neutron activated Au samples were counted with HPGe gamma spectrometers. The $^{198}$Au ($T_{1/2} = 2.69517$ d.) and the $^{198m}$Au ($T_{1/2} = 2.27$ d.) radioisotopes from the $^{197}$Au$(n,\gamma)^{198}$Au nuclear reaction were present in all Au samples irradiated. Our measurements were focused on detection of the $E_\gamma = 332.983$ keV ($I_\gamma = 22.9\%$) and $E_\gamma = 355.684$ keV ($I_\gamma = 87\%$) energy gamma photons emitted due to decay of the $^{196g}$Au ($T_{1/2} = 6.183$ d.) radioisotope that might be produced via dineutron emission in the $^{197}$Au $(n,^2n)^{196}$Au nuclear process. At Atomki the three Au samples (Au-1, Au-2 and Au-3) irradiated in cadmium cover were counted together on the surface of a vertical HPGe detector (Model: GC1520-7935.7, Canberra Industries, Inc., Meiden, Connecticut, USA) with 15% relative efficiency. The arrangements of the measurements are shown in figs. 4 and 5 with number of counts detected/fitted (Y axis) versus energy of gamma rays (X axis).

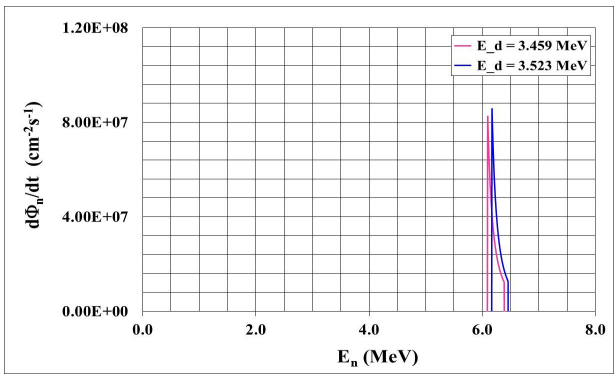

Fig. 3: (Colour on-line) The spectra of the (*d,D*) neutrons for the geometry centres of the foil stacks.

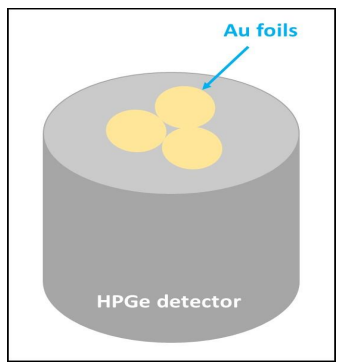

Fig. 4: (Colour on-line) The counting arrangement at the HPGe detector used at Atomki (Debrecen, Hungary).

The gamma lines and corresponding peaks at $E_\gamma$ = 332.983 keV and $E_\gamma$ = 355.684 keV in the gamma spectrum measured for the set of three Au samples are shown in fig. 5.

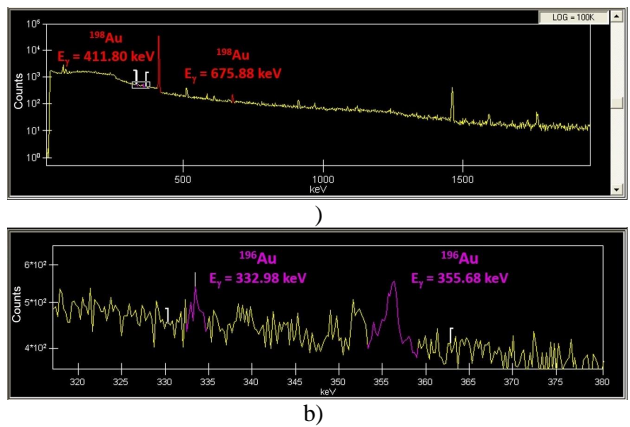

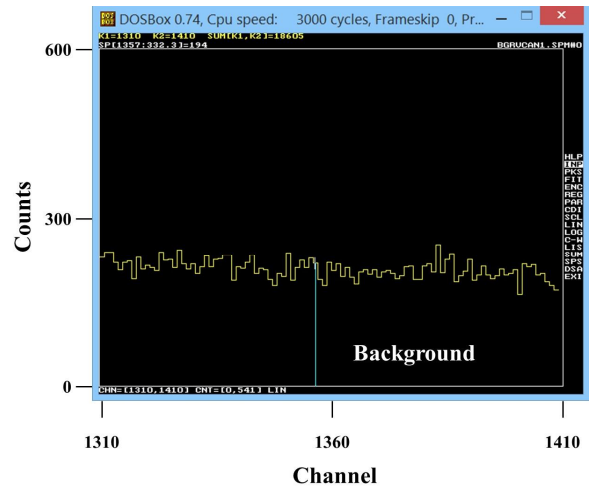

c)

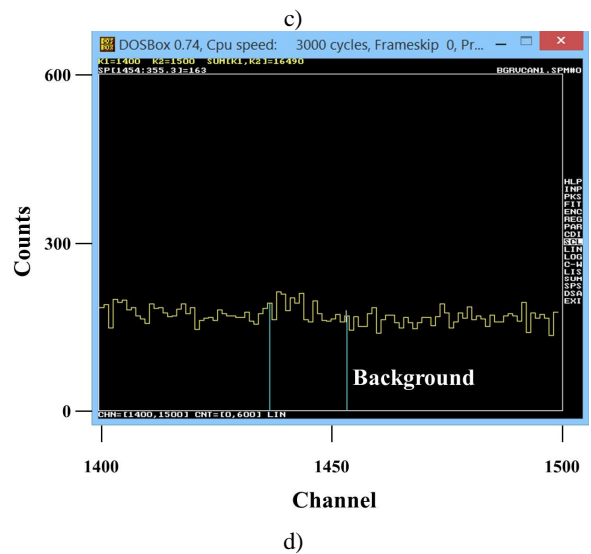

d)

Fig. 5: (Colour on-line) The parts of the gamma spectrum measured at Atomki for $t_{LIVE}$ = 271 482 s for the set of the Au-1 + Au-2 + Au-3 samples: a) - the total spectrum and b) ó ROIs for the $E_\gamma$ = 332.983 keV and the $E_\gamma$ = 355.684 keV gamma lines. The relevant parts of the spectrum measured for $t_{LIVE}$ = 147 864 s with non irradiated samples for the background are also shown in figs c) and d) for the same gamma lines. In fig. b) the ROIs of fitted peaks are presented (violet line).

Corresponding peak areas equal 307±119 and 867±133 counts. The saturation activities 7.8 ± 3.0 and 6.23 ± 0.96 Bq were calculated for $^{196g}$Au using the 332.983 and 355.684 keV gamma lines. This means that these gamma lines/transitions belong to the same decay scheme, namely to that of $^{196g}$Au. The contribution of some extra activity due to 333.82 keV gammas ($I_\gamma$ = 18%) because of $^{198m}$Au decay did not exceed 20 % to the 332.983 keV gamma peak of $^{196g}$Au. More precise estimation is not applicable because the relative statistical uncertainty of the peak area measured for the most intensive gamma line due to $^{198m}$Au decay ($E_\gamma$ = 214.89 keV, $I_\gamma$ = 77.3%) exceeded 100%. The live time and the real time of the counting were $T_{LIVE}$ = 271 482 s and $T_{REAL}$ = 271 640 s, respectively. Therefore, in the case of counting the Au-1, Au-2 and Au-3 samples together at Atomki the statistical significances were 3.7 for the $E_\gamma$ = 332.983 keV gamma peak and 9.4 for the $E_\gamma$ = 355.684 keV gamma peak. Second set of Au foils was counted at INSCU/Department of Nuclear Physics, Taras Shevchenko National University of Kyiv, Ukraine with application of HPGe spectrometer, described in [3]. In the case of counting the Au-I, Au-II and Au-III samples together at INSCU in Kyiv, the best statistical significance achieved was 6.4 for the peak that corresponds to $E_\gamma$ = 355.684 keV gamma photons. After counting the gamma activity of the samples, the $A_{sat}$ saturation activity of $^{196}$Au was calculated.

Then, on the basis of the results of the two irradiations, cross sections were estimated for bound dineutron emission in the outgoing channel of the $^{197}$Au$(n,^2n)^{196g}$Au nuclear process. The cross section averaged for the ($E_{n;min}$, $E_{n;max}$) neutron energy interval was obtained as

$$\overline{\sigma} \equiv \frac{\int_{E_{n;min}}^{E_{n;max}} \phi(E_n)\sigma(E_n)dE_n}{\int_{E_{n;min}}^{E_{n;max}} \phi(E_n)dE_n} = \frac{A_{sat}}{\phi},$$

where $E_n$, $\phi(E_n)$ and $\sigma(E_n)$ are the neutron energy, the energy distribution of the neutron fluence rate and the excitation function of the nuclear reaction, respectively. The measurement and calculation results are summarized in table 1.

Table 1: The results of the counting of the full energy photopeaks of $^{196g}$Au and the estimated cross sections for the dineutron emission in the $^{197}$Au $(n,^2n)^{196g}$Au nuclear process.

| $E_d$, MeV | $E_{n;min}$, MeV | $E_{n;average}$, MeV | $E_{n;max}$, MeV | $E_\gamma$-the gamma peaks used for CS estimation, keV | Cross section (CS), mbarn |
|---|---|---|---|---|---|
| 3.459 | 6.09 | 6.19 | 6.39 | 332.983 355.684 | 0.18 ± 0.06 |
| 3.523 | 6.175 | 6.275 | 6.455 | 355.684 | 0.037 ± 0.008 |

The main contributors for the total uncertainty were as follows:
− positioning of the stacks in front of the $D_2$-gas target, < 10 %;
− limited knowledge on the possible contribution of the neutrons generated by the bombarding deuteron beam impinged on the Ta collimators, the Nb window foil and the W beam stop, < 5 %;
− counting statistics, 22 %;
− uncertainty of the detection efficiencies estimated for the counting geometry used at the HPGe detector at Atomki, 17%.

The partial uncertainties were summed up in quadrature and then the square root of the sum was used for estimation of the total uncertainty.

**Analysis of interfering reactions.** Additional 11 countings were conducted by us at INSCU to check whether a half-life estimate for $^{196g}$Au did not vary due to possible cosmic bursts. Based on measurement results, no indications were found within experimental uncertainties that $^{196g}$Au activity was induced by a cosmogenic activation. This activity dropped below the detection level within 1 month after the irradiation was completed.

As it was mentioned above, Au foils at Atomki were additionally checked for impurities and besides Au (99.9 w%) only Hg (0.1 w%) isotopes with natural isotopic abundance and $^{209}$Bi (<0.01 w%) were detected. These mean that only the $^{196}$Hg $(n,X)^{196}$Au nuclear processes could be contributing to the production of $^{196}$Au with $(d,D)$ neutrons. According to the TENDL-2019 version of the TENDL nuclear data library [6], the evaluated cross sections of the $^{196}$Hg $(n,X)^{196}$Au reaction are $\sigma$ = 9.6 mb at $E_n$ = 6.0 MeV and $\sigma$ = 24.5 mb at $E_n$ = 6.8 MeV, respectively. Therefore, by our estimates the contribution of the $^{196}$Hg $(n,X)^{196}$Au nuclear interference to the $^{196g}$Au production was less than 1%.

In six months after irradiation with $(d,D)$ neutrons, the Au-III sample was irradiated with $(d,T)$ neutrons to check out Hg impurity by detection of the $^{197}$Hg induced activity due to the $^{198}$Hg $(n,2n)$ nuclear reaction with cross section ~ 2b and the $^{196}$Hg $(n,\gamma)$ reaction with cross section ~ 1.3 mb. No 134 keV ($I_\gamma$=33.5%) gamma peak due to decay of $^{198}$Hg was present in the instrumental gamma spectrum and only reaction products from $(n,2n)$, $(n,p)$, $(n,\alpha)$ and $(n,\gamma)$ reactions on $^{197}$Au were reliably detected. This means our Au samples do not contain any detectable amount of Hg.

Also, considering possible tritium build-up from the $d(d,T)p$ nuclear reaction in irradiation of samples with $(d,D)$ neutrons [7], our choice of a $D_2$-gas target was not occasional with idea to minimize the effect of $(d,T)$ neutrons on fast neutron activation of Au foils through the $(n,2n)$ nuclear reaction. In order to exclude a likeliness of the observed gamma peaks of 332.983 keV and 355.983 keV energies owing to $(d,T)$ neutrons interaction with Au nuclei, we assumed that the activity of $^{196}$Au was apparently induced by $(d,T)$ neutrons only. The cross section of the $D(d,p)T$ nuclear reaction is 84.7 mb at $E_d$ = 3.459 MeV deuteron energy. Therefore, neutrons with $E_n$ = 20.2 MeV energy are produced via the $T(d,n)^4$He reaction, too, in the $D_2$-gas [7]. The cross section of the $T(d,n)^4$He reaction equals 88.8 mb at $E_d$ = 3.459 MeV deuteron energy. Thus, the activation of the Au foils by the $(d,T)$ neutrons via the $^{197}$Au$(n,2n)^{196g}$Au reaction with cross section of 1.3 b has to be taken into account. By the end of the irradiation the density of the produced T atoms was $n_T$ = 2.3•10$^{11}$ cm$^{-3}$ in the gas cell assuming their uniform distribution in the $D_2$-gas. The estimated maximum fluence rate of the $E_n$ = 20.2 MeV energy neutrons at the geometry centre of the stack of the (Au-I + Au-II + Au-III) foils was $\phi_h \approx$ 8.9·10$^{-3}$ cm$^{-2}$s$^{-1}$. This neutron fluence rate results in about 6·10$^{-4}$ % contribution of the $^{197}$Au($n,2n$)$^{196}$Au reaction to the measured net areas of the gamma peaks observed at 332.983 keV and 355.983 keV energies. These mean that $I_d \approx 700$ $\mu$A deuteron beam would be needed for producing the amount of the measured $^{196g}$Au activity and the corresponding measured peak areas if they were induced exclusively by $E_n$ = 20.2 MeV energy ($d,T$) neutrons. However, $I_d \approx 50$ $\mu$A is the maximum beam current that can be extracted from the MGC-20E cyclotron of Atomki at $E_d$ = 3.5 MeV deuteron energy. Inanalogously, similar estimate was done for $^{196}$Hg ($n,p$)$^{196g}$Au nuclear reaction with $\phi_h$ (see above), 0.01 w% of Hg in our samples, 0.15% of $^{196}$Hg isotope in natural abundance of Hg and 20 mb cross section for 20.2 MeV neutrons [6] in the input channel. This estimate gives about 1.4·10$^{-11}$ % contribution to $^{196g}$Au production in our samples of gold via this nuclear reaction channel.

**Discussion.** One could also assume, that we may be facing with two neutron halo configuration nucleus, i.e. $^{196}$Au + $n$ + $n$, like for some light nuclei. By definition, for such nuclei the two neutrons separation energy ($S_{2n}$) must be less than 1 MeV, but for $^{198}$Au $S_{2n}$ equals 14.585 MeV, so it is certainly not the case. In [3] we provided the first evidence of a bound dineutron appearing as the two-neutron nucleus in the outgoing channel of the $^{159}$Tb ($n,2n$)$^{158}$Tb nuclear reaction based on prediction by A. Migdal in [2]. Our experimental results and some estimates in [3] were done for a configuration of the dineutron located in a close proximity to $^{158}$Tb ó the nucleus of prolate shape with deformation parameter $\beta_{Tb}$ = 0.271. The current study presents reliable observation of a bound dineutron near the surface of $^{196}$Au ó the nucleus of oblate shape with deformation parameter $\beta_{Au}$ = - 0.125. Meanwhile, one of the most valuable theoretical expressions in [2], providing the formulae for calculation of energy of the single particle level, includes radius of the nucleus $R$ as a parameter and was derived for a spherical shape of a nucleus without deformation. Thus, rather than the only single particle state for each of a nucleus of spherical shape there is the possibility for deformed nuclei of prolate and oblate shapes to have not only one, but rather two locations for a bound dineutron to be settled in. Therefore, taking into account our experimental data from both experiments, we can not exclude the possibility of dineutron formation not only in the direction of Z-axis, but also near the surface of corresponding nucleus in the direction of X or Y axis. Based on two cross-section estimates, which are among the major results presented in this research, one can assume that the dependence of cross section versus energy for the $^{197}$Au($n,2n$)$^{196}$Au nuclear process might start from zero in the vicinity of the energy $E= E_{th;(n,2n)} - B_{dn}$, increase and, upon reaching the maximum within [$E$, 6.17] MeV, subsequently decrease after 6.17 MeV and approaching zero value near the $E_{th;(n,2n)}$ energy, with which the ($n,2n$) nuclear reaction channel opens up. Here $B_{dn}$ is the binding energy of the dineutron, and according to current interval estimate, $B_{dn}$ stays within [2.2-2.8] MeV [1]. In other words, the cross section versus energy dependence has a resonant character, as predicted in [2]. At the same time if the two single particle levels could be formed and available for population by a bound dineutron, then this resonant behaviour of the $^{197}$Au($n,2n$)$^{196}$Au nuclear reaction may have fine structure of this reaction cross section. Last attribute might serve as a possible basis for explanation of significant reduction in the $^{197}$Au($n,2n$)$^{196}$Au reaction cross section with 85 keV shift in neutron spectrum as described above. Furthermore, if the two single particle levels of different energies can accept 2+1 neutrons, then there is potential to generate the trineutron ($^3n$), as it was discussed in [1] based on available experimental data from the $^{159}$Tb ($p,3n$)$^{157}$Dy nuclear reaction. At this stage we can not conclude whether such unique configuration could be a bound nucleus, consisting of the three neutrons, or just a combination like $^2n + n$, trapped together at single particles levels near the surface of deformed nuclei. Most probably, this concern may be addressed if corresponding decay data could be obtained [1].

Thus, statistically significant observation of a bound dineutron in the $^{197}$Au($n,2n$)$^{196}$Au nuclear reaction, and in addition the true conclusion about a manifestation of dineutron escape in the $^{159}$Tb($n,2n$)$^{158}$Tb nuclear reaction in [3], both demonstrated, that A. Migdaløs prediction about possible formation and existence of a bound dineutron in the outgoing channel of nuclear reaction is absolutely correct nowadays even it was made 45 plus years ago with some obsolete values like n-n scattering lengths within and beyond the nucleus, etc. As confirmed by our experimental data, the cross section versus energy dependence for the $^{197}$Au($n,2n$)$^{196}$Au nuclear reaction has a resonant character, to some extent similar to the pumping between levels in a laser system. By analogy, letøs call this phenomenon as õthe dineutron pumpingö. Then thorough study of such cross section vs energy dependence with application of VdG accelerators, targeted to precisely identify a maximum position and corresponding cross section value, could enable us to deeply understand between which levels in and beyond of residual nucleus this dineutron pumping takes place. Also, we can note the mass interval, for which dineutron generation seems to be possible now is as follows: 159-197. This interval is in a full agreement with what was predicted in [1] and [3].

**Summary. -** Results of this study shed some light on the process of reliably generating such unique particles as bound dineutrons, which were considered as non-existent for very long time since the first mentioning by Colby and Little in [8]. While the theoretical description of dineutron formation in the vicinity of much heavier nucleus is for now not that much comprehensive, our statistically significant experimental results, exceeding the 5 criteria, confirmed observation of the dineutron, which since now is not just hypothesized to exist, but is certainly existing. We also were capable of deriving cross-section estimates 180 ± 60 and 37 ± 8 $\mu$b for the two energy intervals of incident neutrons [6.09 ÷ 6.39] and [6.175 ÷ 6.455] MeV, respectively. Thus, based on our experimental facts, this research opens up a new exciting direction in nuclear physics, appealing for much more efforts to explore new low energy physics. Then our observations allow us to raise a reasonably justified question about introduction of a new nuclear reaction channel ($n,^2n$) into available nuclear libraries. Moreover, this initiative is in a full conformance with the following statement from [6]: *"TENDL takes a rather extreme point of view here: every nuclear reaction process which is expected to take place in reality should be present in a nuclear data library, measured or not measured. In short, that means, all projectiles, all nuclides, all reaction channels and all energies"*.

∗∗∗

The authors acknowledge the European Regional Development Fund and Hungary in the frame of the project GINOP-2.2.1-15-2016-00012 for partial financial support of this research. Separate thanks to crews of the MGC-20E cyclotron at Atomki for their contribution to the reliable operation of this facility for irradiation experiments. Thank Attila Csík (Atomki) for his performing the SIMS analysis of the Au samples.